\documentclass[twocolumn]{aastex63}
\usepackage{amsmath}
\usepackage{amssymb}

\definecolor{purple}{rgb}{0.5,0,0.5}
\definecolor{darkgreen}{rgb}{0.1,0.6,0.1}
\definecolor{orange}{rgb}{1,0.6,0}

\newcommand{\brvs}{Brunt-V\"ais\"al\"a}

\submitjournal{ApJ}
\shorttitle{Convective Boundary Mixing}
\shortauthors{Jermyn et al.}

\begin{document}

\title{Convective Penetration in Early-Type Stars}

\correspondingauthor{Adam S. Jermyn}
\email{adamjermyn@gmail.com}

\author[0000-0001-5048-9973]{Adam S. Jermyn}
\affiliation{Center for Computational Astrophysics, Flatiron Institute, New York, NY 10010, USA}

\author[0000-0002-3433-4733]{Evan H. Anders}
\affiliation{CIERA, Northwestern University, Evanston, IL 60201, USA}

\author[0000-0002-7635-9728]{Daniel Lecoanet}
\affiliation{CIERA, Northwestern University, Evanston, IL 60201, USA}
\affiliation{Department of Engineering Sciences and Applied Mathematics, Northwestern University, Evanston IL 60208, USA}

\author[0000-0001-5048-9973]{Matteo Cantiello}
\affiliation{Center for Computational Astrophysics, Flatiron Institute, New York, NY 10010, USA}
\affiliation{Department of Astrophysical Sciences, Princeton University, Princeton, NJ 08544, USA}

\begin{abstract}
Observations indicate that the convective cores of stars must ingest a substantial amount of material from the overlying radiative zone, but the extent of this mixing and the mechanism which causes it remain uncertain.
Recently,~\citet{2021arXiv211011356A} developed a theory of convective penetration and calibrated it with 3D numerical hydrodynamics simulations.
Here we employ that theory to predict the extent of convective boundary mixing in early-type main-sequence stars.
We find that convective penetration produces enough mixing to explain core masses inferred from asteroseismology and eclipsing binary studies, and matches observed trends in mass and age.
While there are remaining uncertainties in the theory, this agreement suggests that most convective boundary mixing in early-type main-sequence stars arises from convective penetration.
Finally, we provide a fitting formula for the extent of core convective penetration for main-sequence stars in the mass range from $1.1-60 M_\odot$.
\end{abstract}


\section{Introduction}\label{sec:intro}

Observations tell us that stars have more mixing near the boundaries of convection zones than theoretical models predict.
For instance asteroseismology of B stars reveals that their convective cores are larger than predicted by 1D evolutionary models by tens of percent~\citep{2021NatAs...5..715P}, and a similar conclusion comes from modelling of early type eclipsing binary systems~\citep{2019ApJ...876..134C}.
Likewise observations of lithium depletion in F stars hint at extra mixing beneath surface convection zones~\citep{1995ApJ...446..203B}, as do measurements of the location of the base of the solar convection zone from helioseismology~\citep[][Sct. 7.2.1]{2016LRSP...13....2B}.

The extent and profile of this extra mixing matters critically for our understanding of stellar evolution.
For instance, the lifetimes and end points of massive star evolution depend directly on how much fresh fuel is mixed into their cores~\citep{2011A&A...530A.115B,2017RSOS....470192S,2019ApJ...887...53F}.
Likewise the efficiency of processes like dredge up~\citep{1986PASP...98.1066S} and the position of the Red Giant Branch luminosity bump~\citep{2018ApJ...859..156K} depend sensitively on convective boundary mixing (CBM).

These challenges and opportunities have motivated a great deal of theoretical work identifying and characterizing mechanisms for CBM.
These include convective overshooting and penetration~\citep{zahn1991, brummell_etal_2002, 10.1093/mnras/stz047,shaviv_salpeter_1973, herwig2000} as well as mixing by internal gravity waves launched by convection~\citep{1991ApJ...377..268G,2017ApJ...848L...1R}.

Recently,~\citet{2021arXiv211011356A} proposed a theoretical model of convective penetration and calibrated this to 3D hydrodynamical simulations of Boussinesq convection.
Here we study the implications of the resulting calibrated model in early-type stars.
We begin in Section~\ref{sec:theory} by reviewing the theory of convective overshoot and penetration and presenting the model.
We then explain how we implement this model in Section~\ref{sec:methods}.
We show predictions of this model for the cores of early-type stars in Section~\ref{sec:results} and compare these with observations.
These predictions were produced by \emph{post-processing} stellar evolution models run with no convective penetration, to better show the instantaneous extent of this mixing rather than its cumulative effect over evolutionary time-scales.
We find that convective penetration produces enough mixing to explain core masses inferred from observations, and matches the observed trends in mass and age.
We then discuss uncertainties and future work in Section~\ref{sec:discussion}, and conclude in Section~\ref{sec:conclusions}.

\begin{figure}
\centering
\includegraphics[width=0.48\textwidth]{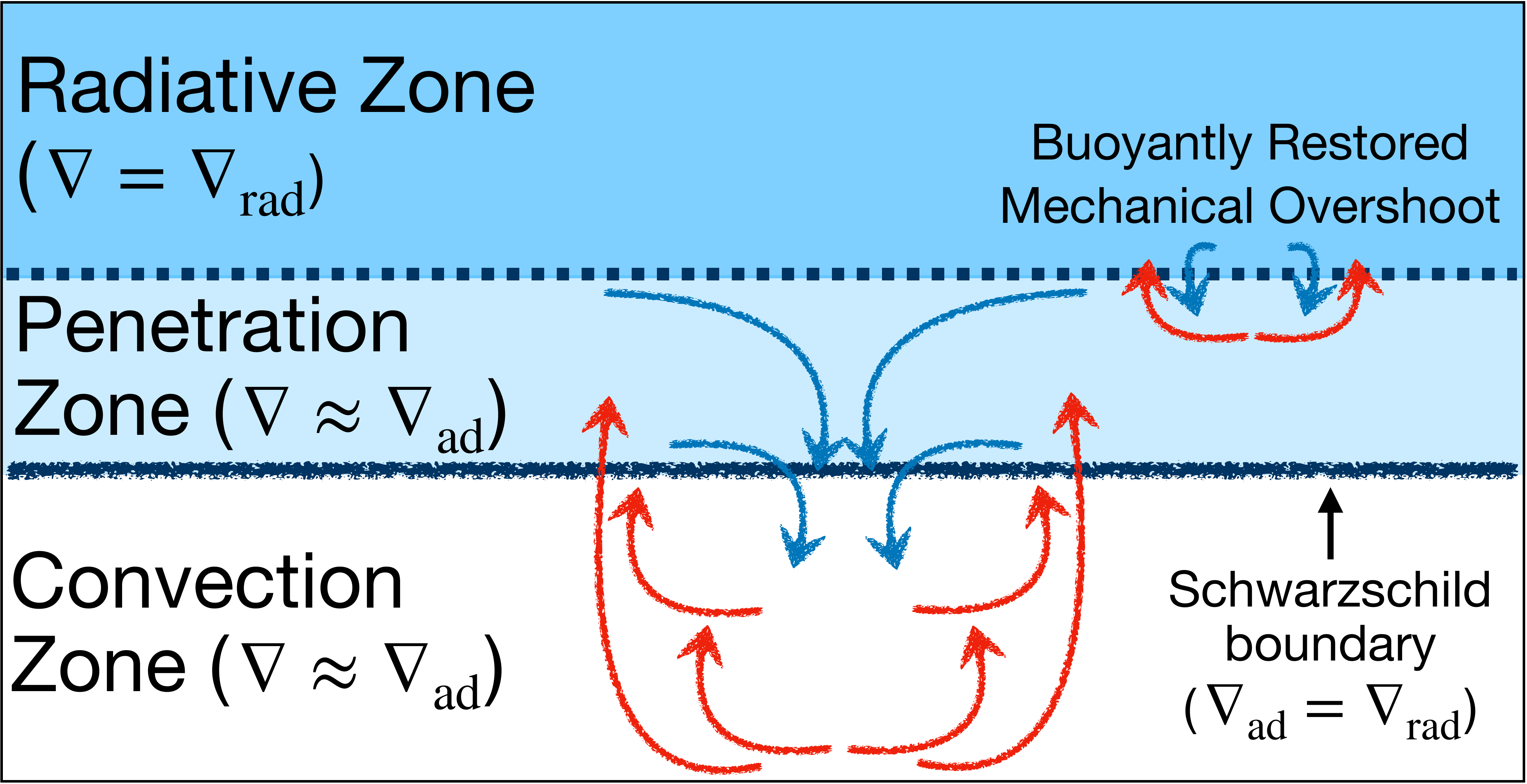}
\caption{The structure of a convective boundary as found by~\citet{2021arXiv211011356A} is shown schematically. The top of the domain is a radiative zone, where $\nabla=\nabla_{\rm rad}$. Below this is the penetration zone, which has $\nabla \approx \nabla_{\rm ad}$. Motions in the penetration zone which hit the boundary with the radiative zone experience a large buoyant restoring force, and so this mechanical overshooting only reaches a small distance into the radiative zone. Beneath the penetration zone is the convection zone, which is slightly superadiabatic. Because the penetration zone is nearly adiabatic and hence nearly neutrally buoyant, motions which start in the convection zone experience a very weak restoring force in traversing the penetration zone. This allows the penetration zone to remain thermally well-mixed.}
\label{fig:schema1}
\end{figure}

\section{Theory}\label{sec:theory}

In 1D stellar evolution calculations, the boundary of a convection zone (CZ) is typically set at the point where the \brvs\ frequency $N$ vanishes.
At this point local fluid motions feel no net gravitational acceleration.
In a system with uniform composition this is the Schwarzschild boundary, where
\begin{align}
	\nabla_{\rm rad} = \nabla_{\rm ad}.
	\label{eq:sc}
\end{align}
Here
\begin{align}
	\nabla_{\rm ad} \equiv \left.\frac{\partial \ln T}{\partial \ln P}\right|_s
\end{align}
is the adiabatic temperature gradient and $s$ is the entropy{\bf,
\begin{align}
	\nabla_{\rm rad} = \frac{3\kappa\rho L P}{64 \pi G M \sigma T^4}
\end{align}}
is the radiative temperature gradient {\bf in the diffusive approximation}, $\kappa$ is the opacity, $\rho$ is the density, $L$ is the luminosity, $P$ is the pressure, $T$ is the temperature, $G$ is the gravitational constant, $M$ is the mass below the point of interest, and $\sigma$ is the Stefan-Boltzmann constant.
In the presence of composition gradients, the boundary is instead where the Ledoux criterion is marginally satisfied, such that
\begin{align}
	\nabla_{\rm rad} = \nabla_{\rm ad} + \nabla_\mu
	\label{eq:sl}
\end{align}
and for an ideal gas
\begin{align}
	\nabla_\mu \equiv \frac{d\ln \mu}{d\ln P},
\end{align}
where $\mu$ is the mean molecular weight.

Convective motions carry momentum, however, and so can extend past the surface of neutral buoyancy, producing overshoot.
After moving beyond the convective boundary, eddies experience a buoyant restoring acceleration $\sim h N^2$, where $N$ is evaluated in the stable layer.
{\bf Here
\begin{align}
	h \equiv -\frac{dr}{d\ln P}
\end{align}
is the pressure scale height.}

An eddy with velocity $v_c$ can then move past the convective boundary for a time $\Delta t \sim v_c / h N^2$, during which time it moves a distance
\begin{align}
	\Delta r \sim v_c \Delta t \sim \frac{v_c^2}{h N^2},
	\label{eq:dr_ov}
\end{align}
where $N$ is evaluated in the stable layer{\bf, typically a distance $\sim h$ away from the convective boundary}, $v_c$ is the typical velocity of convective motions{\bf, and $r$ is the radial coordinate}.

The overshooting given by equation~\eqref{eq:dr_ov} is called mechanical overshoot and is often very small, typically $\Delta r/h \la 10^{-6}$ for convective cores on the main sequence.
More sophisticated models based on the same momentum picture provide larger values~\citep{10.1093/mnras/stz047}, but these are generally still much too small to match the extent of convective boundary mixing inferred from observations~\citep{2021NatAs...5..715P}.

In addition to carrying momentum, though, convective motions also carry heat.
This allows convective overshooting to alter the thermal structure of the star near the point of neutral buoyancy and ultimately form a \emph{layer} of neutral buoyancy, known as a penetration zone (PZ).
The PZ is an adiabat, with temperature gradient $\nabla \approx \nabla_{\rm ad}$, and so material can flow freely through it unhindered by buoyancy.
Overshooting still occurs, but begins at the outer edge of the PZ, rather than the edge of the Schwarzschild-unstable region (Figure~\ref{fig:schema1}).
This process is known as convective penetration, and can result in more extended mixing layers than mechanical overshooting.

Building on~\citet{zahn1991} and~\citet{roxburgh1989},~\citet{2021arXiv211011356A} developed a theory of convective penetration to predict the extent of the PZ and calibrated that theory to 3D cartesian hydrodynamical simulations of Boussinesq convection\footnote{These simulations were conducted over a thermal time-scale ($\sim 10^4$ turnover times) because that is the time over which a statistically stationary penetration zone develops.}
This theory is based on an exact rewriting of the time-and-spatially averaged energy equation, accounting for all energy fluxes, sources, and sinks.
Several of the terms arising can be extracted from 1D stellar models, while others must be calibrated from numerical simulations.
All told, the extent of the penetration zone is then given by the integral constraint
\begin{equation}
-\frac{\int_{\rm{PZ}} L_{\rm{conv}}\,dr}{\int_{\rm{CZ}} L_{\rm{conv}}\,dr} + f \xi \frac{V_{\rm{PZ}}}{V_{\rm{CZ}}} = (1 - f),
\label{eq:mesa_eqn}
\end{equation}
where
\begin{equation}
f = 0.86\qquad\rm{and}\qquad
\xi = 0.6
\label{eqn:f_xi_estimates}
\end{equation}
are calibrated from numerical simulations, $L_{\rm conv}$ is the convective luminosity, $V_{\rm PZ}$ is the volume of the penetration zone, and $V_{\rm CZ}$ is the volume of the convection zone.
The convective luminosity in the PZ is given in thermal equilibrium by
\begin{align}
	L_{\rm conv, PZ} = -L \left(\frac{\nabla_{\rm ad} - \nabla_{\rm rad}}{\nabla_{\rm rad}}\right).
	\label{eq:conv_PZ}
\end{align}
We emphasize that the calibration in equations~\eqref{eqn:f_xi_estimates} and our theory more generally are only valid in the Boussinesq limit, and more general models must be used in strongly stratified systems~\citep{roxburgh1989}.

While~\citet{2021arXiv211011356A} did not include the effects of composition gradients, we do not expect these to change the equilibrium extent of the penetration zone.
Overshooting motions still occur when there is a composition gradient at the convective boundary.
These motions serve to erase the composition gradient by mixing material in the overshooting region.
Over time this allows overshooting to erase the composition gradient and build a penetration zone in its wake.

Note that in terms of compositional mixing convective penetration most naturally matches the step overshooting prescription used in stellar evolution software instruments, because motions in the penetration zone are nearly as fast as those in the convection zone, and the boundary between the penetration zone and the overlying stable layer is sharp~\citep{2021arXiv211011356A}.
In addition, however, convective penetration mixes entropy to produce an adiabatic layer.
This entropy mixing is not typically implemented as part of step overshooting and is not standard in any stellar evolution software instrument of which we are aware, though it has been implemented in individual calculations~\citep{2021NatAs...5..715P}.

\section{Methods}\label{sec:methods}

We calculated stellar evolutionary tracks for stars ranging from $1.1-60 M_\odot$ using revision 15140 of the Modules for Experiments in Stellar Astrophysics software instrument
\citep[MESA][]{Paxton2011, Paxton2013, Paxton2015, Paxton2018, Paxton2019}.
Details on the MESA microphysics inputs are provided in Appendix~\ref{appen:mesa}.
Our models use the Milky Way metallicity $Z=0.014$ and determine convective stability using the Ledoux criterion.
We further employ convective premixing~\citep[][Section 5.2]{Paxton2019} to more accurately model the evolution of the convective boundary in the presence of composition gradients.

Using equations~\eqref{eq:mesa_eqn} and~\eqref{eq:conv_PZ} we calculated the extent of the penetration zone just outside the convective core.
We emphasize that this calculation is done in post-processing: all stellar evolution calculations were performed with no convective penetration or additional boundary mixing.
This likely makes a difference to the evolution of a model: as convective penetration grows the convective core it changes the temperature profile, which in turn can change the extent of convective penetration, particularly if the edge of the core gets pushed into a region with a very different opacity profile.

Our motivation in doing this calculation in post-processing was just to understand the rough scale and trends involved in convective penetration. To enable precision comparisons with observations, future work should perform this calculation self-consistently with stellar evolution.

Along similar lines, because our models have a composition gradient in the vicinity of the convective boundary, and because that gradient would be erased in the presence of a penetration zone, we calculate $\nabla_{\rm rad}$ and $\nabla_{\rm ad}$ for purposes of equations~\eqref{eq:mesa_eqn} and~\eqref{eq:conv_PZ} using the \emph{core composition} everywhere.
We likewise use these constant-composition temperature gradients to determine the location of the nominal convective boundary (i.e. the base of the penetration zone), which may differ from that reported by MESA due to e.g. convective premixing and compositional gradients.

Our approach of using the core composition for the penetration zone is not quite right, because penetrative motions mix material of a slightly different composition into the core, but the amount of this mixing is small enough that to first order we may use the core composition without any penetration to determine the extent of the penetration zone.
Improving on this approximation again requires performing convective penetration self-consistently with stellar evolution, which we leave for the future.

\section{Results}\label{sec:results}

\subsection{Radial Extent}

\begin{figure*}
\centering
\begin{minipage}{0.49\textwidth}
\includegraphics[width=\textwidth]{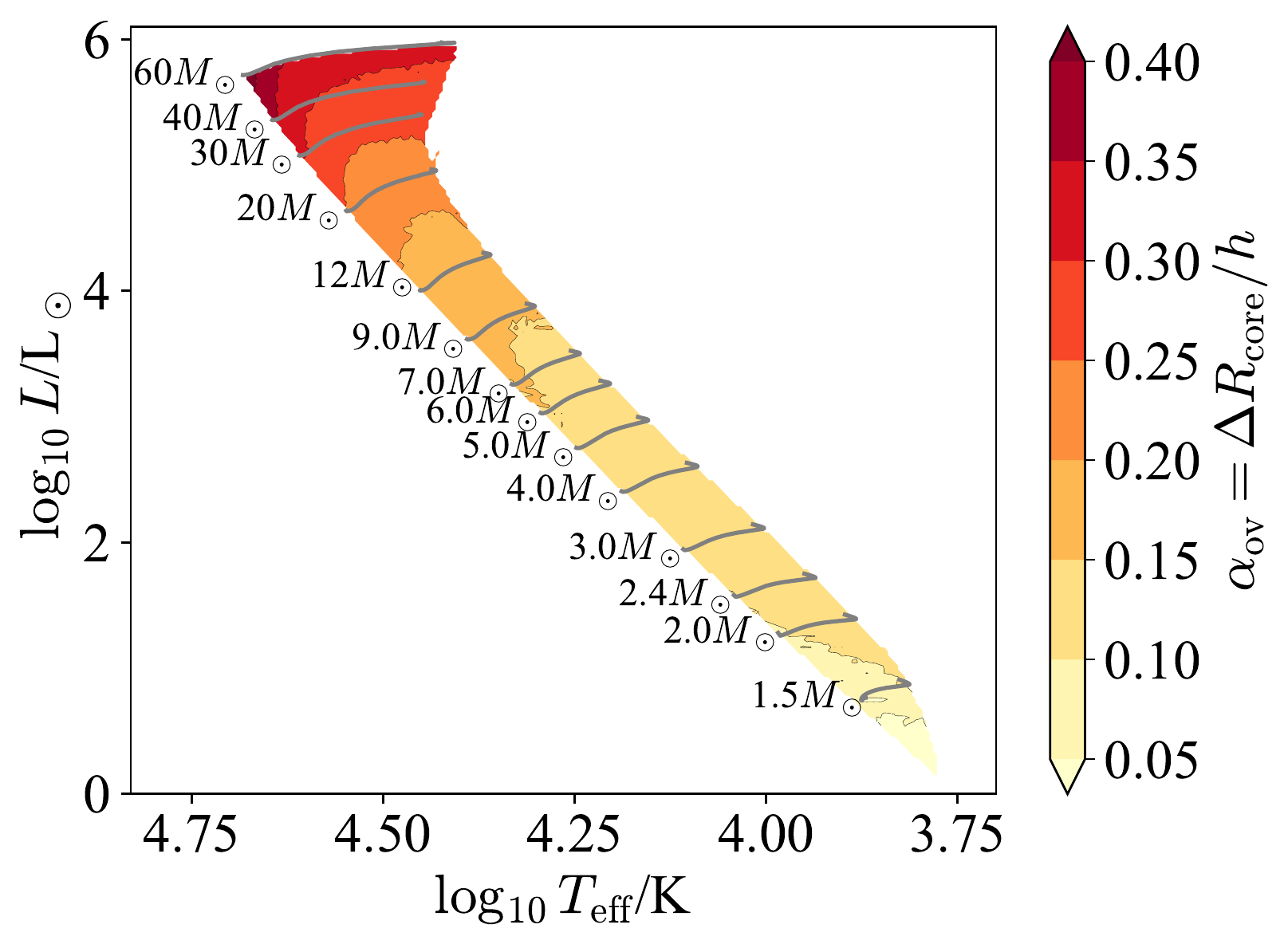}
\end{minipage}
\hfill
\begin{minipage}{0.49\textwidth}
\includegraphics[width=\textwidth]{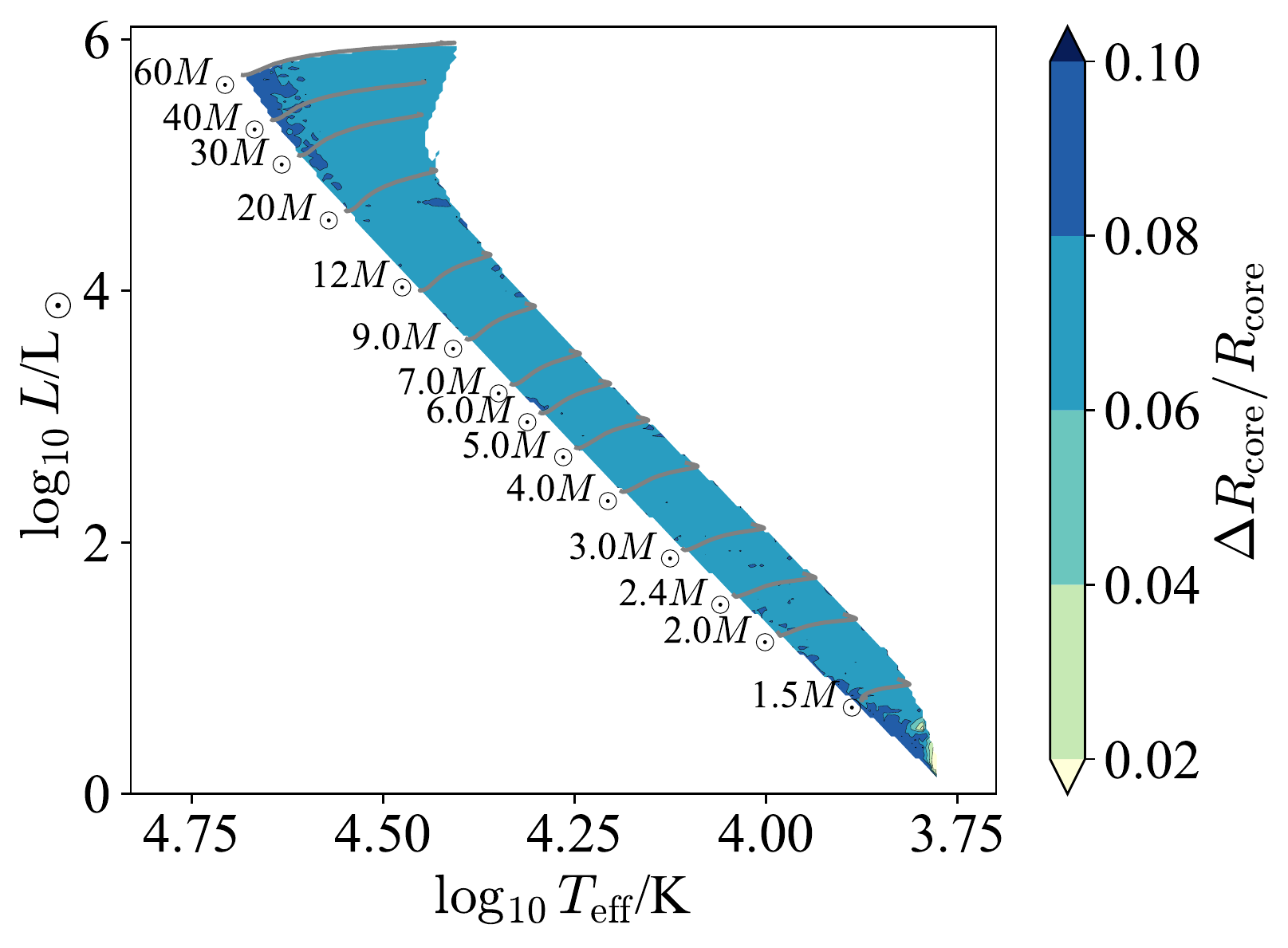}
\end{minipage}

\caption{The radial extent of the penetration zone is shown on a Hertzsprung-Russel diagram in terms of $\log T_{\rm eff}$ and $\log L$ for stellar models ranging from $1.1-60 M_\odot$ with Milky Way metallicity $Z=0.014$. (Left) Normalized to the pressure scale height at the nominal convective boundary reported by MESA. (Right) Normalized to the radius at the same. Values below the lower edge of the scale occur at masses $M \la 1.5 M_\odot$ as the core convection zone is just forming.}
\label{fig:dr}
\end{figure*}

Figure~\ref{fig:dr} shows the radial extent of the penetration zone on a Hertzsprung-Russel diagram, normalized to the pressure scale height $h$ (left) and the radius (right), both at the nominal boundary of the convective core.
The quantity reported on the left is equivalent to the commonly-used step overshooting parameter $\alpha_{\rm ov}$, which gives the step overshooting distance as a fraction of a pressure scale-height.

We see clear trends in mass for $\alpha_{\rm ov}$ (left).
Higher mass stars show much larger penetration zones relative to the pressure scale height.
We see mixing of $\alpha_{\rm ov} \sim 0.1$ for A-stars, $\alpha_{\rm ov}\sim 0.3$ for B-stars, and up to $\alpha_{\rm ov} \sim 0.45$ for the most massive O-stars.

\begin{figure}
\centering
\includegraphics[width=0.48\textwidth]{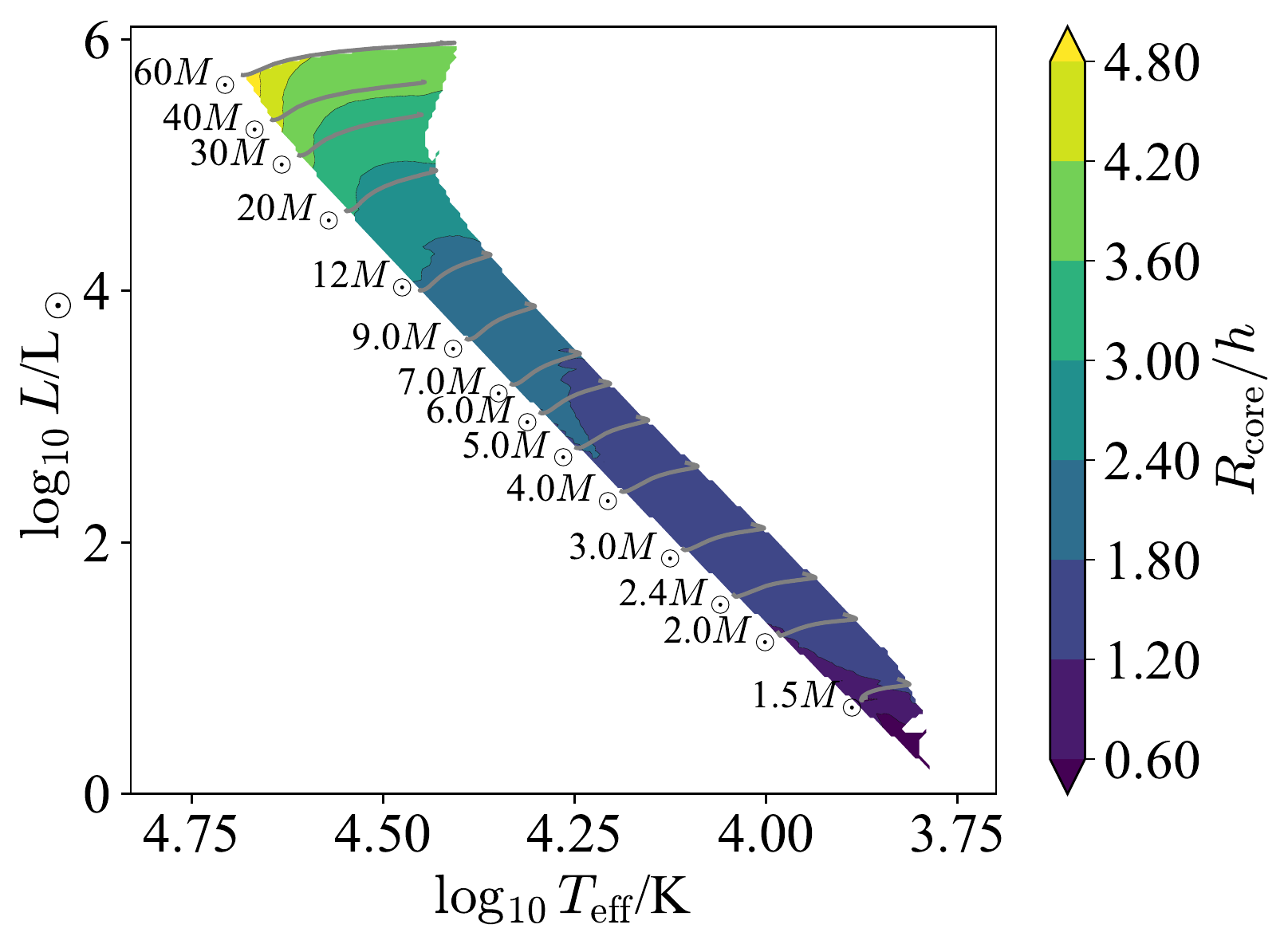}
\caption{The ratio $R/h$ at the outer boundary of the nominal convective core is shown on a Hertzsprung-Russel diagram in terms of $\log T_{\rm eff}$ and $\log L$ for stellar models ranging from $1.1-60 M_\odot$ with Milky Way metallicity $Z=0.014$. Note that even though this ratio becomes large, the Boussinesq approximation remains valid over much of the domain, as evidenced by the density ratio in Figure~\ref{fig:scale_heights}.}
\label{fig:R_div_h}
\end{figure}

By contrast, when we measure the penetration zone in terms of $\Delta R_{\rm core}/R_{\rm core}$ (Figure~\ref{fig:dr}, right) the trend with mass almost vanishes.
This tells us that the trend in $\alpha_{\rm ov}$ is being driven almost entirely by the changing volume of the convection zone.
That is, the ratio $R_{\rm core}/h$ is varying (Figure~\ref{fig:R_div_h}), and that in turn causes the extent of the penetration zone to vary via the term $f\xi V_{\rm PZ}/V_{\rm CZ}$ appearing in equation~\eqref{eq:mesa_eqn}.

\begin{figure}
\centering
\includegraphics[width=0.48\textwidth]{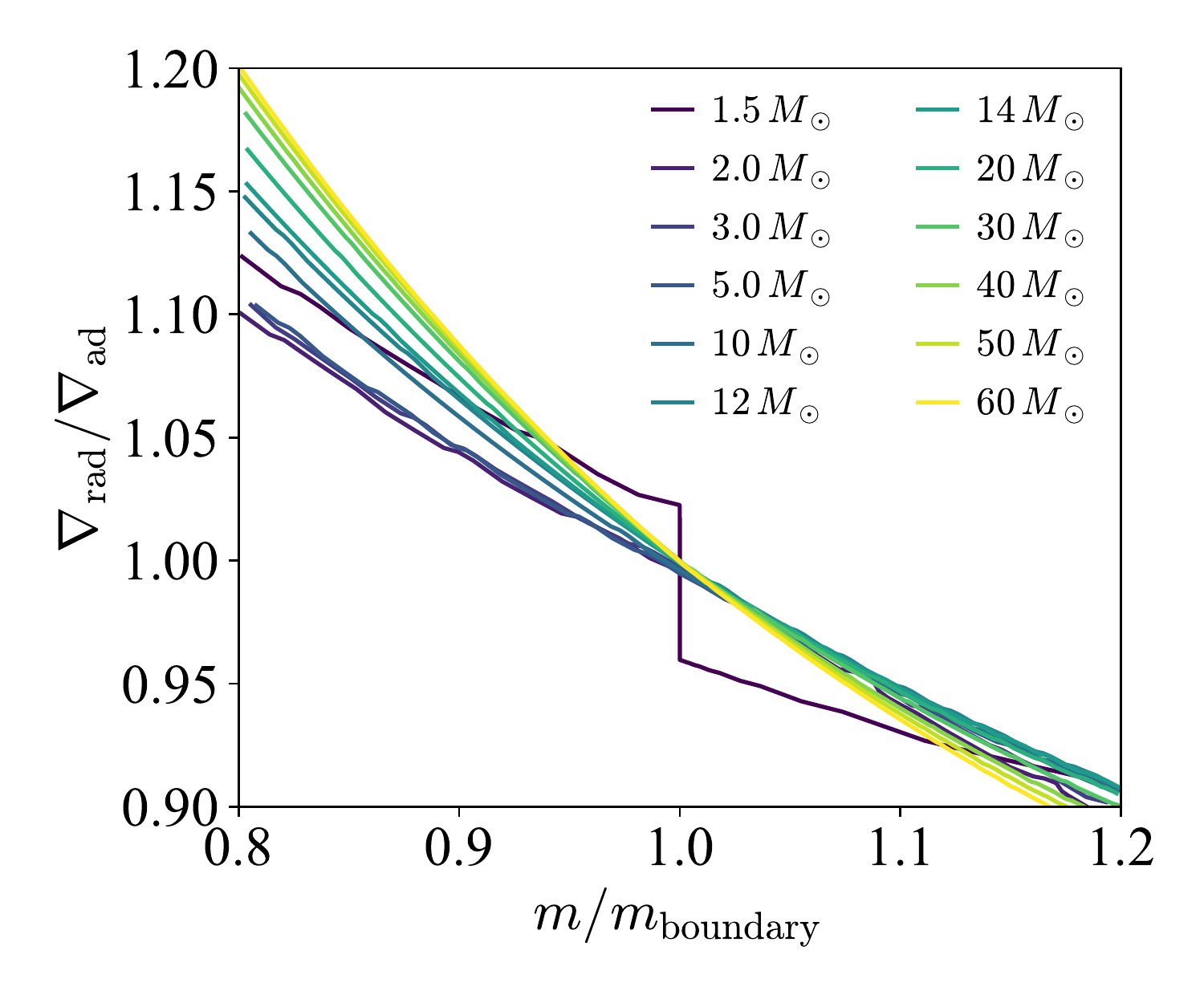}
\caption{Profiles of $\nabla_{\rm rad}/\nabla_{\rm ad}$ are shown for selected mid-main-sequence stellar models as a function of mass coordinate normalized to the nominal convective boundary.
These profiles are calculated assuming a fixed composition matching that of the core, and so are the ones entering into equation~\eqref{eq:conv_PZ}.
They are \emph{not} the profiles actually used in the stellar evolution.}
\label{fig:P_param}
\end{figure}

The fact that the volume of the convection zone sets the trends in Figure~\ref{fig:dr} is supported by Figure~\ref{fig:P_param}, which shows mid-main-sequence (central hydrogen abundance $X_c=0.36$) profiles of $\nabla_{\rm rad}/\nabla_{\rm ad}$ as a function of mass coordinate normalized to the nominal convective boundary.
These profiles are calculated assuming a fixed composition matching that of the core, and so are the ones entering into equation~\eqref{eq:conv_PZ}.
They are \emph{not} the profiles actually used in the stellar evolution.

Except for the $1.5 M_\odot$ model, the profiles are extremely similar, meaning that the ratios of convective to total luminosity in equation~\eqref{eq:conv_PZ} are similar, and hence so are the ratio of luminosity integrals in equation~\eqref{eq:mesa_eqn}.
This leaves only the volume of the convection zone to set the scale of the penetration zone.

The $1.5 M_\odot$ model, and those at lower masses too, show a somewhat different profile, which is driven by a composition discontinuity at the convective boundary.
Even though we use a fixed composition in computing these profiles, we use the actual density profiles from our stellar models, and those can exhibit a discontinuity at the nominal convective boundary if there is a composition discontinuity at the same.
Despite this jump, we see just a small difference in $\Delta R/R_{\rm core}$ in Figure~\ref{fig:dr}, and so believe that even at low masses the trend is dominated by the volume of the convection zone rather than any composition effects.

\begin{figure}
\centering
\includegraphics[width=0.48\textwidth]{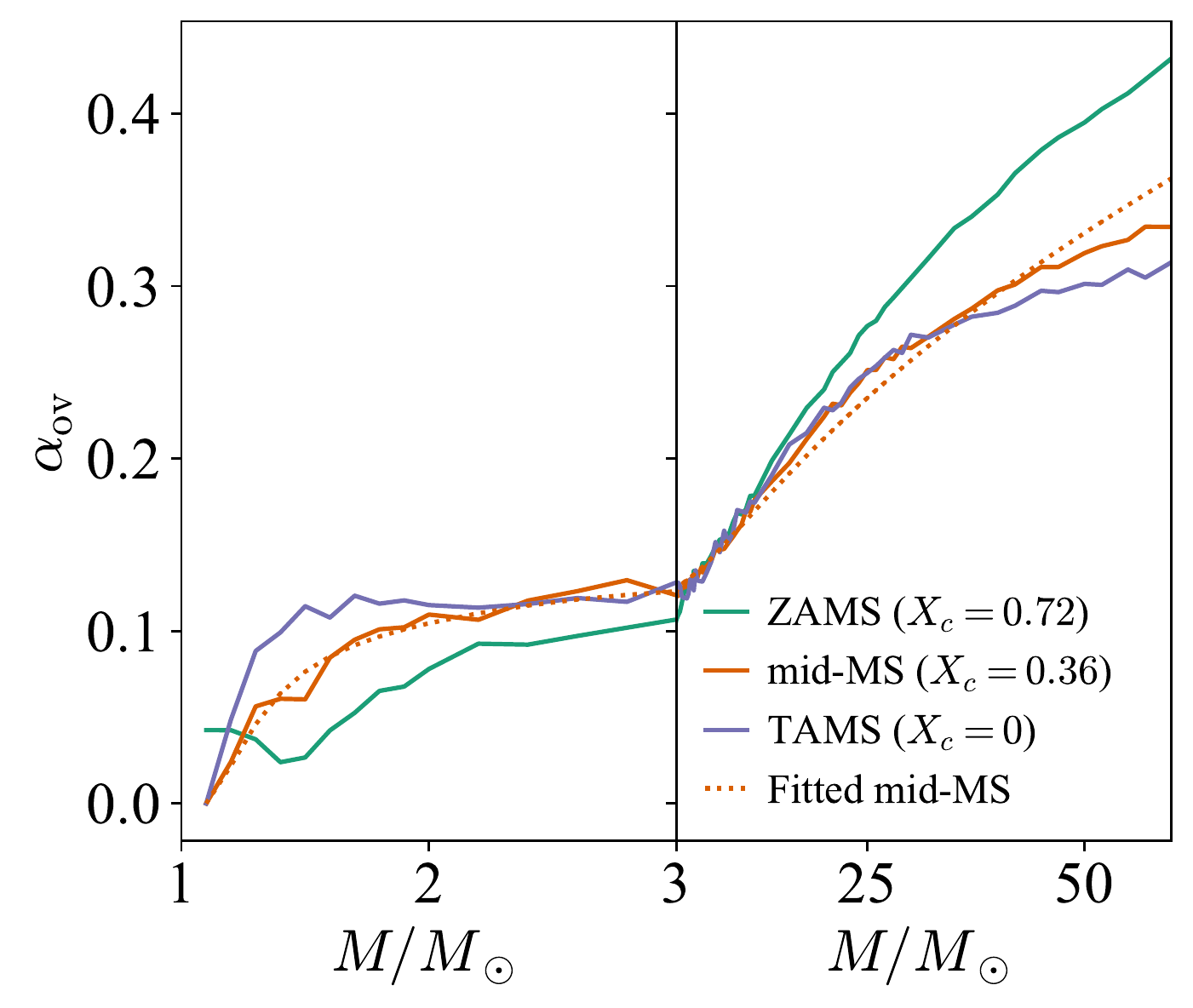}
\caption{The $\alpha_{\rm ov}$ step overshooting parameter needed to match the predicted extent of the penetration zone is shown as a function of mass for three different points in time, the ZAMS, the TAMS, and the midpoint between them as measured by central $^{1}{\rm H}$ abundance. The dashed line shows the fit to the mid-aged models given by equation~\eqref{eq:form}.}
\label{fig:fov_mass}
\end{figure}

We likewise see clear trends in age for $\alpha_{\rm ov}$.
At the high mass end ($M \ga 30 M_\odot$) the penetration zone shrinks by 20-30\% with age from the Zero-Age Main Sequence (ZAMS) to the Terminal-Age Main Sequence (TAMS), while at the low mass end it actually expands slightly with age ($M \la 3 M_\odot$).
This may be seen more clearly in Figure~\ref{fig:fov_mass}, which shows our predicted $\alpha_{\rm ov}$ as a function of mass at the ZAMS, the TAMS, and the halfway point between these (mid-MS).
As with mass, this trend almost entirely disappears for $\Delta R_{\rm core}/R_{\rm core}$, again because the volume of the convection zone dominates the trends of $\alpha_{\rm ov}$.

The residual trends we see in $\Delta R_{\rm core}/R_{\rm core}$ are principally at the low-mass end ($M \la 2 M_\odot$).
There, the core mass increases with age as the nuclear luminosity of the CNO cycle increases, increasing $\nabla_{\rm rad}$ and changing its profile~\citep[][see figure 3.6]{2010aste.book.....A}.
Later, once the CNO cycle is dominant, $\nabla_{\rm rad}$ falls and the core mass declines\footnote{This decline is responsible for the jump in $\nabla_{\rm rad}/\nabla_{\rm ad}$ in our $1.5 M_\odot$ model in Figure~\ref{fig:P_param}.}.
These effects are also all functions of mass, and so give rise to both a mass and age dependence in $\Delta R_{\rm core}/R_{\rm core}$.

Figure~\ref{fig:fov_mass} also shows good agreement with the trends inferred by~\citet{2019ApJ...876..134C} from eclipsing binary stars, at least over the range from $1-3M_\odot$.
Both our predictions and their inferences yield $\alpha_{\rm ov}$ increasing rapidly with increasing mass, up to around $2 M_\odot$, and then a much weaker trend beyond that.
We further predict that this trend of increasing $\alpha_{\rm ov}$ with increasing mass continues beyond the edge of their data ($\approx 3 M_\odot$) up to quite large masses ($60 M_\odot$).
We emphasize that the trends we see in $\alpha_{\rm ov}$ are dominated by the volume of the nominal convection zone and not by the temperature gradient profiles, as the latter are quite uniform in mass and age.

We additionally have reasonable quantitative agreement on the magnitude of $\alpha_{\rm ov}$: they infer a plateau around $2 M_\odot$ of $f_{\rm ov} \approx 0.015$ and say that the equivalent $\alpha_{\rm ov}$ is $11.4$ times larger, giving $\alpha_{\rm ov} \approx 0.17$, which is a little higher than our prediction of $\alpha_{\rm ov} \approx 0.12$ for the mid-MS but well within the uncertainties of their inferences and of e.g. converting between step-overshooting ($\alpha_{\rm ov}$) and exponential overshooting ($f_{\rm ov}$).

{\bf By contrast~\citet{2010A&A...515A..87D} inferred $\alpha_{\rm ov} \approx 0.17$ for the solar-like star HD~203608, which is quite a bit larger than our predicted $\alpha_{\rm ov} \sim 0.05$ for low-mass ZAMS stars, though that object is also at sub-solar metallicity, which could change the extent of mixing.}

While we recommend using equation~\eqref{eq:mesa_eqn} to predict the extent of the penetration zone in stellar models, there are instances where a fitting formula is more convenient.
We find a good fit to our mid-main-sequence results with
\begin{align}
	\alpha_{\rm ov} = \tanh(m-1.1)\sqrt{m-1.1}\frac{a + b m^2 + c m^3 + d m^5}{e + m^5}
	\label{eq:form}
\end{align}
where $m \equiv M_\star/M_\odot$ and
\begin{align}
	a &= 2.47109,\\
	b &= -1.19087,\\
	c &= 0.724183,\\
	d &= 0.0470249,\\
\intertext{and}
	e &= 0.560757
\end{align}
are fitting coefficients.
Equation~\eqref{eq:form} produces a relative error of at most 21\% and an absolute error of at most $0.02$.
These errors are smaller than the difference between $\alpha_{\rm ov}$ on the ZAMS and TAMS.

\subsection{Mass Extent}

We can also view these predictions in mass-coordinate.
The left panel of Figure~\ref{fig:dm} shows the mass of the penetration zone relative to the mass of the nominal convective core.
We see that the additional fractional mass which penetration mixes into the core declines with increasing mass, though does so gradually, adding at most $30\%$ to the mass of the core and so extending the lifetime of the star by a similar amount.

\begin{figure}
\centering
\includegraphics[width=0.48\textwidth]{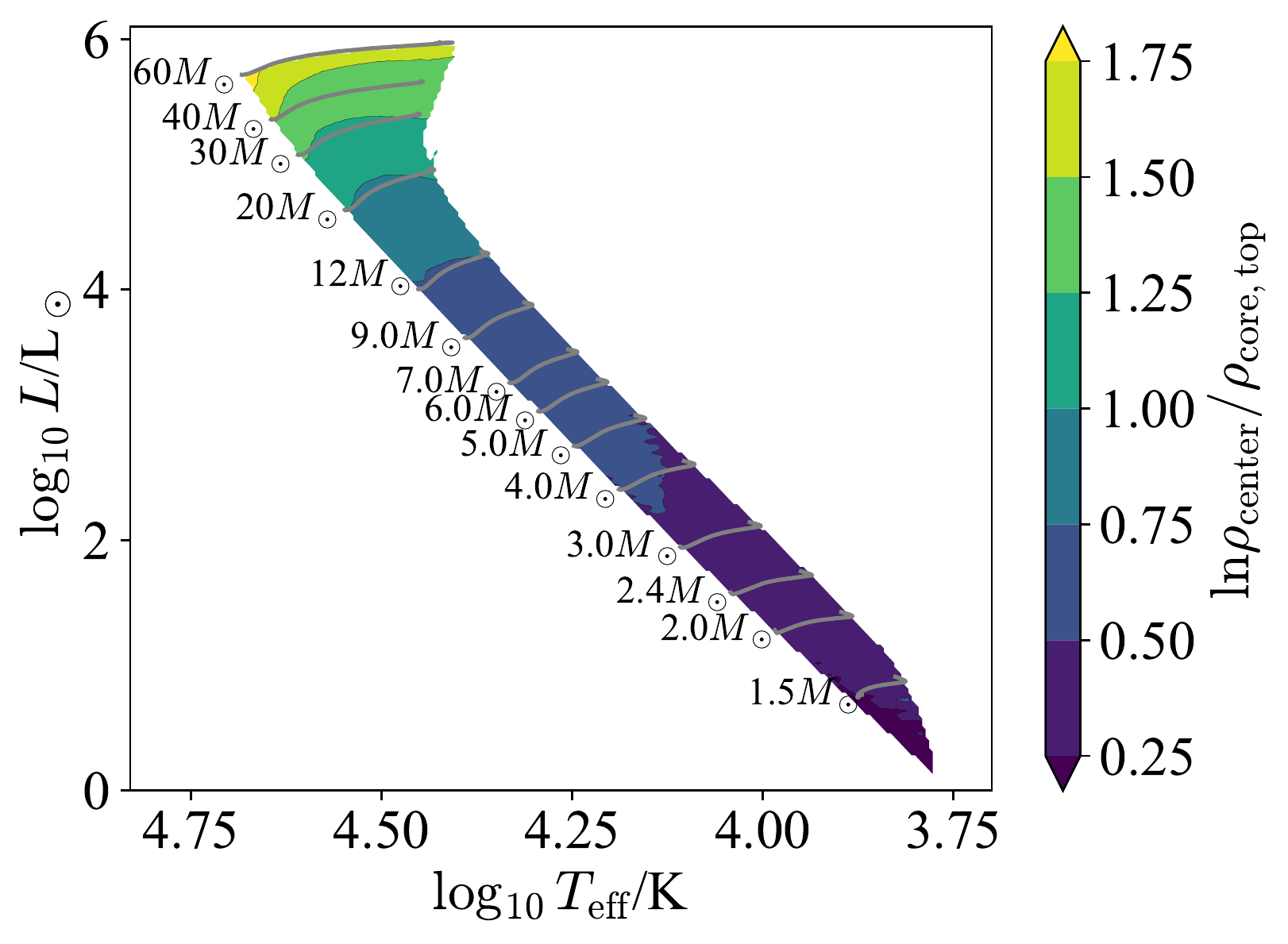}
\caption{The number of pressure scale heights in the nominal convective core is shown on a Hertzsprung-Russel diagram in terms of $\log T_{\rm eff}$ and $\log L$ for stellar models ranging from $1.1-60 M_\odot$ with Milky Way metallicity $Z=0.014$.}
\label{fig:scale_heights}
\end{figure}

This trend could be due in part to our formulation (equation~\ref{eq:mesa_eqn}) and calibrating simulations neglecting density stratification.
As the mass of the star increases the core grows and the number of scale heights it contains increases (Figure~\ref{fig:scale_heights}).
Thus a fixed relative increase in the \emph{radial} extent of the core, as seen in Figure~\ref{fig:dr}, translates to a smaller increase in the mass extent because of the increasing density contrast between the penetration zone and the center of the star.
We intend to examine the effects of density stratification in future work.

\begin{figure*}
\centering
\begin{minipage}{0.49\textwidth}
\includegraphics[width=\textwidth]{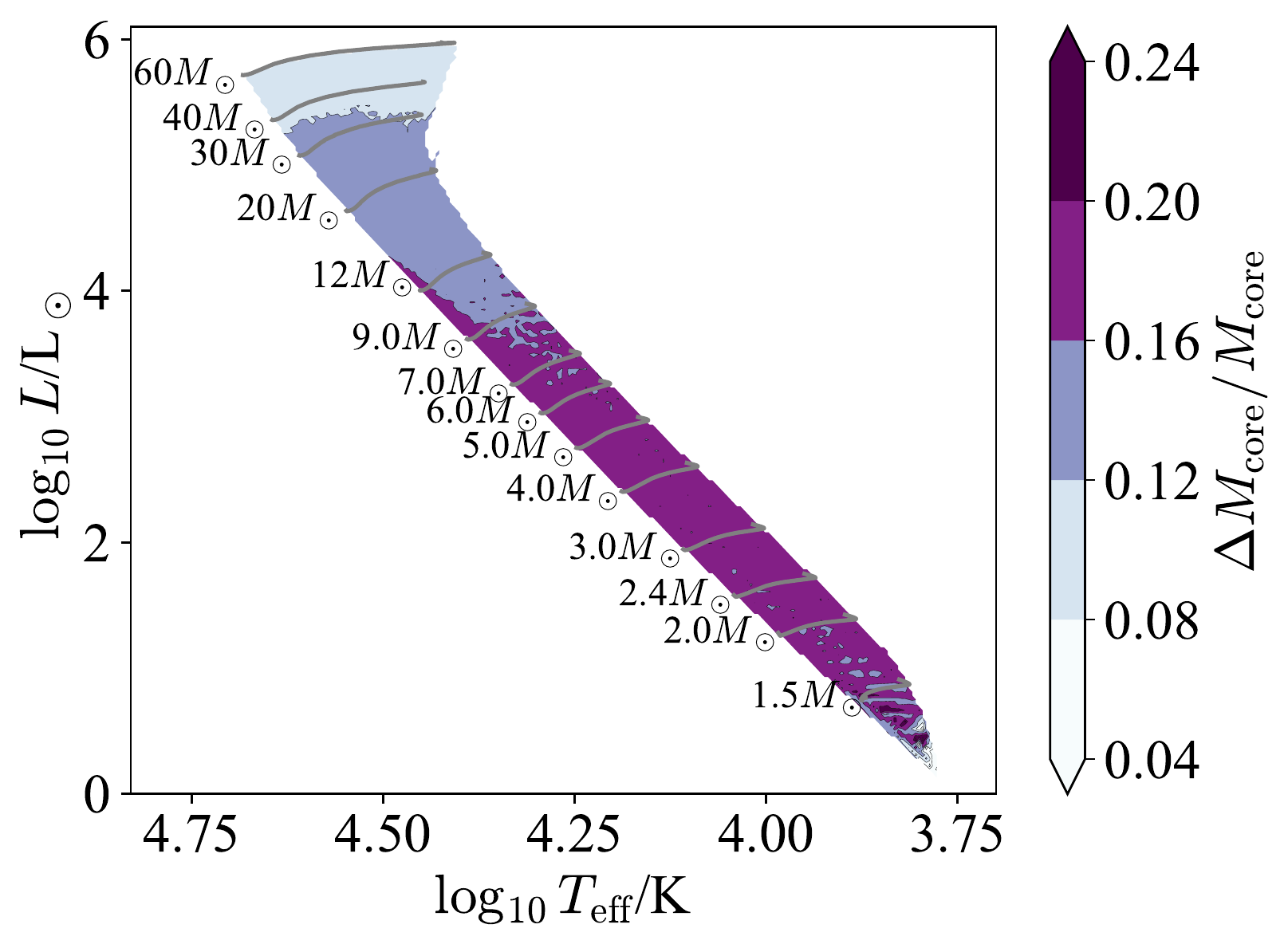}
\end{minipage}
\hfill
\begin{minipage}{0.49\textwidth}
\includegraphics[width=\textwidth]{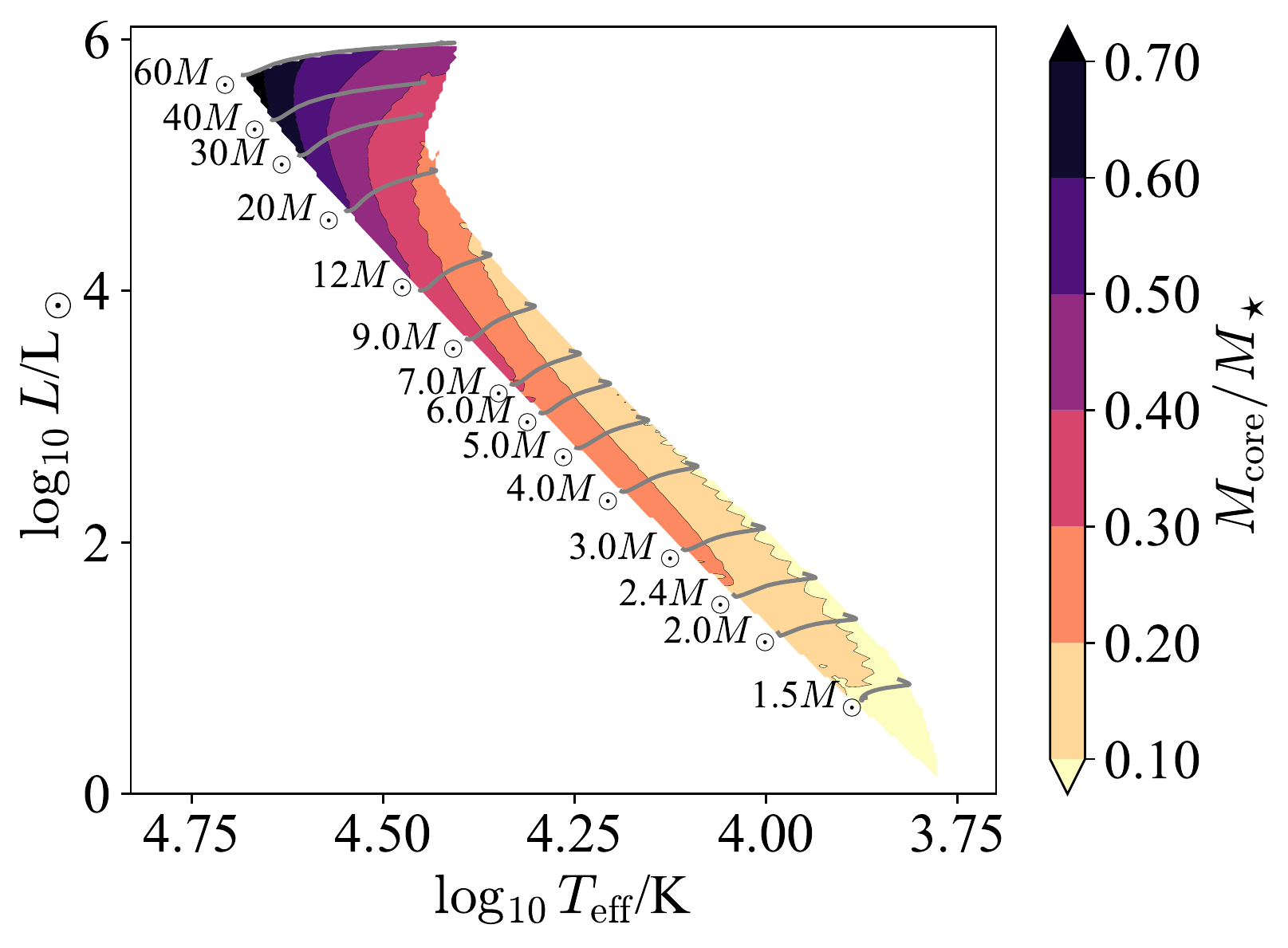}
\end{minipage}

\caption{(Left) The mass of the penetration zone is shown on a Hertzsprung-Russel diagram in terms of $\log T_{\rm eff}$ and $\log L$ for stellar models ranging from $1.1-60 M_\odot$ with Milky Way metallicity $Z=0.014$, normalized to the mass coordinate at the nominal convective boundary reported by MESA.  Values below the lower edge of the scale occur at masses $M \la 1.5 M_\odot$ as the core convection zone is just forming. (Right) The fractional mass of the star inside the outer edge of the penetration zone is shown on the same diagram for the same models.}
\label{fig:dm}
\end{figure*}

\begin{figure}
\centering
\includegraphics[width=0.48\textwidth]{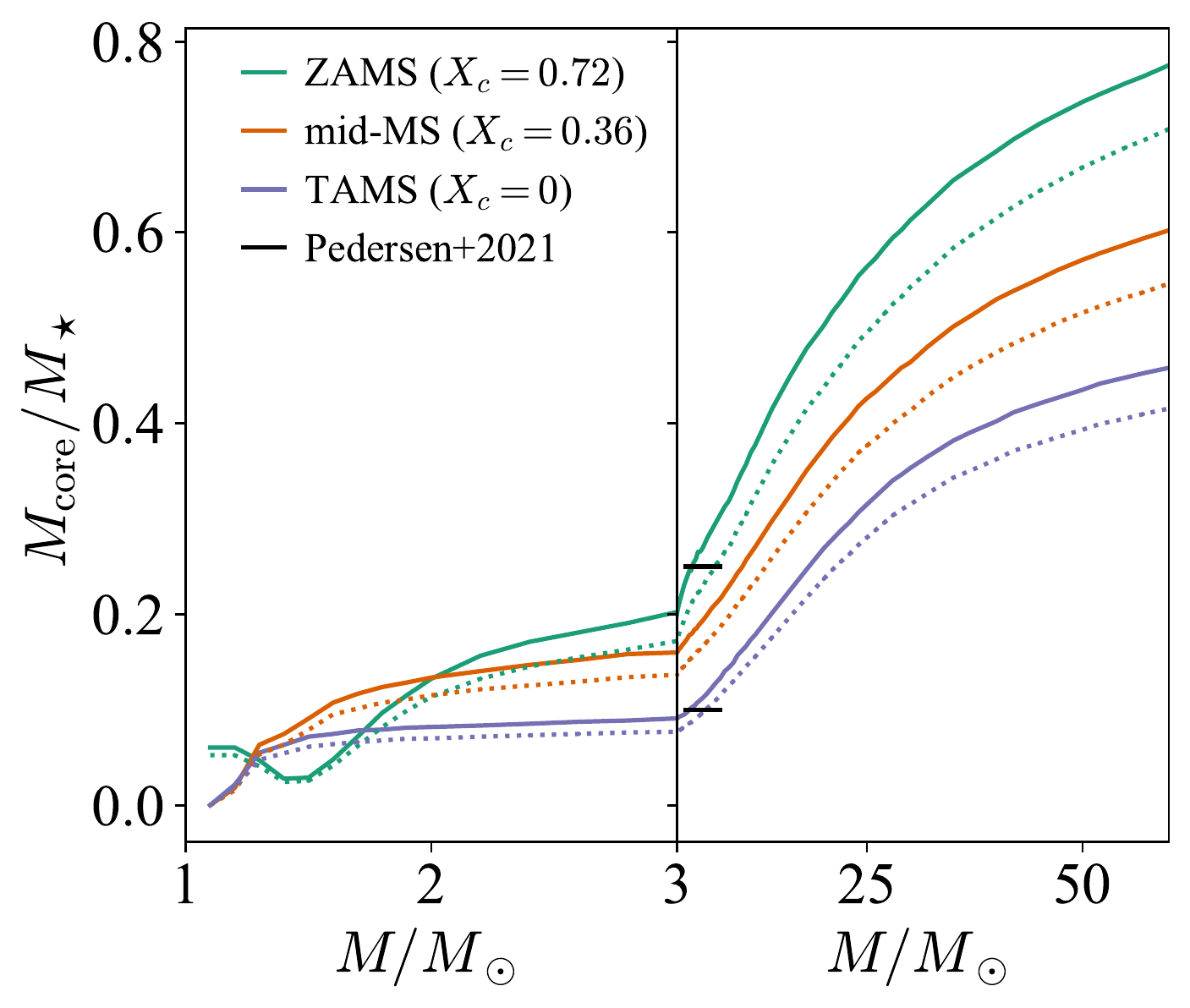}
\caption{The fractional mass of the star inside the outer edge of the penetration zone is shown as a function of mass for three different points in time, the ZAMS, the TAMS, and the midpoint between them as measured by central $^{1}{\rm H}$ abundance. Dashed lines indicate the nominal core mass with no penetration zone and solid lines show the core mass including the penetration zone. Light horizontal lines indicate the approximate measurements on the ZAMS and TAMS from~\citet{2021NatAs...5..715P}.}
\label{fig:mc_mass}
\end{figure}

Note that the fraction of a star's total mass in the core including the penetration zone shows somewhat different trends.
The right panel of Figure~\ref{fig:dm} shows the fraction of the total mass which resides in the well-mixed core.
The relative core mass increases monotonically with total mass and decreases monotonically with age.
This may also be seen in Figure~\ref{fig:mc_mass}, which shows our predicted relative core mass as a function of total mass at the ZAMS, the TAMS, and the halfway point between these.
The nominal core mass increases rapidly with stellar mass and decreases rapidly with age, and these effects suffice to overcome trends in the relative mass of the penetration zone.

With Figure~\ref{fig:mc_mass} we can also compare our predictions to the trends inferred from asteroseismology of B stars.
\citet{2021NatAs...5..715P} find approximately $M_{\rm core}/M_\star \approx 0.25$ on the ZAMS and $\approx 0.1$ on the TAMS for stars ranging from $4-8M_\odot$ (see their Figure 6).
This is in quantitative agreement with our models which include convective penetration (solid lines) and significantly greater than if we exclude the penetration zone (dashed lines).
\citet{2021NatAs...5..715P} also see a weak trend on the TAMS towards increasing fractional core mass with increasing stellar mass, of order an increase of $\Delta(M_{\rm core}/M_\star)\approx 0.1$  from $4-8M_\odot$, which agrees roughly with our predictions.

\section{Discussion}\label{sec:discussion}

There are several uncertainties which require further study to provide a complete theory of convective penetration.
For instance~\citet{2021arXiv211011356A} only examined cartesian domains in the Boussinesq limit.
For stellar cores a spherical geometry is more appropriate, and may change the obtained values of $f$ and $\xi$ in equation~\eqref{eq:mesa_eqn}, though we do not expect this to change the orders of magnitude or the trends we see.

More importantly, at higher masses we see increasing density stratification in stellar cores, and convective envelopes are even more strongly density-stratified, so extending this work to higher mass main-sequence stars or Red Giants dredge up requires moving beyond the Boussinesq approximation.
\citet{roxburgh1989} suggests a way to adapt this theory for stratified systems.
This approach was applied to stellar models by neglecting dissipation, producing an upper bound on the amount of penetration~\citep{1992A&A...266..291R,1993JRASC..87Q.185D,2004PASP..116..997V,VandenBerg_2006}, but so far as we are aware this has not yet been done including dissipation and calibrating with simulations.
We know, however, that the existing formalism of~\citet{2021arXiv211011356A} is insufficient because equation~\eqref{eq:mesa_eqn}, applied to the Red Giant Branch, predicts at points that the whole star is well-mixed, which is clearly ruled out by observations.

Additionally the effects of rotation and magnetism on convective penetration remain poorly understood, though there has been recent progress on the former~\citep{2018sf2a.conf..135A}.
Incorporating these effects may prove important to explain the diversity of mixing profiles inferred from observations~\citep{2021NatAs...5..715P}.

\section{Conclusions}\label{sec:conclusions}

We have applied a recent theory of convective penetration by~\citet{2021arXiv211011356A} to predict the extent of convective core boundary mixing in early-type stars owing to this mechanism.
We find good agreement both in magnitude and in trends with inferred core masses from asteroseismology and eclipsing binary studies.
This suggests that convective penetration may be responsible for most of the convective boundary mixing which occurs in these stars, and motivates the inclusion of this process in stellar evolution software instruments.
Importantly, while existing instruments support including the effects of convective penetration on compositional mixing via a step overshoot prescription, they do not generally feature a corresponding adiabatic layer, which matters for making precise predictions of asteroseismic signatures~\citep{2021NatAs...5..715P}.

\acknowledgments

The Flatiron Institute is supported by the Simons Foundation.
This research was supported in part by the National Science Foundation under Grant No. PHY-1748958.
We are grateful to May Gade Pedersen and Ben Brown for productive conversations on these topics.


\software{
\texttt{MESA} \citep[][\url{http://mesa.sourceforge.net}]{Paxton2011,Paxton2013,Paxton2015,Paxton2018,Paxton2019},
\texttt{MESASDK} 20190830 \citep{mesasdk_macos,mesasdk_linux},
\texttt{matplotlib} \citep{hunter_2007_aa}, 
\texttt{NumPy} \citep{der_walt_2011_aa}}

\clearpage

\appendix

\section{MESA} \label{appen:mesa}

The MESA EOS is a blend of the OPAL \citep{Rogers2002}, SCVH
\citep{Saumon1995}, FreeEOS \citep{Irwin2004}, HELM \citep{Timmes2000},
and PC \citep{Potekhin2010} EOSes.

Radiative opacities are primarily from OPAL \citep{Iglesias1993,
Iglesias1996}, with low-temperature data from \citet{Ferguson2005}
and the high-temperature, Compton-scattering dominated regime by
\citet{Buchler1976}.  Electron conduction opacities are from
\citet{Cassisi2007}.

Nuclear reaction rates are from JINA REACLIB \citep{Cyburt2010} plus
additional tabulated weak reaction rates \citet{Fuller1985, Oda1994,
Langanke2000}.
Screening is included via the prescription of \citet{Chugunov2007}.
Thermal neutrino loss rates are from \citet{Itoh1996}.

Models were constructed on the pre-main sequence with $Y=0.024+2Z$ and $X=1-Y-Z$ and evolved from there.

Convective penetration is calculated by re-casting equation~\eqref{eq:mesa_eqn} in the form
\begin{align}
\int_{\rm{PZ}} -L_{\rm{conv}} +  f \xi 4\pi r^2 F_{\rm avg}\,dr = (1-f)V_{\rm CZ} F_{\rm avg},
\label{eq:pen}
\end{align}
where
\begin{align}
	F_{\rm avg} \equiv \frac{1}{V_{\rm CZ}}\int_{\rm CZ} L_{\rm conv} dr.
\end{align}
and where $L_{\rm conv}$ is evaluated in the penetration zone using equation~\eqref{eq:conv_PZ}.
We then integrate the left-hand side of equation~\eqref{eq:pen} away from the nominal convective boundary returned by MESA until we obtain the right-hand side.

\section{Data Availability}

The plotting scripts used in this work may be found in this GitHub \href{https://github.com/adamjermyn/cbm_ra_plots}{repository} in the commit with short-sha 2fc2b48. The inlists and run scripts used in producing the HR diagrams in this work are available in this other GitHub \href{https://github.com/adamjermyn/conv_trends}{repository} on the main branch in the commit with short-sha 964e07d. The data those scripts produced are available in~\citet{jermyn_adam_2022_5878965}.

\bibliography{refs}
\bibliographystyle{aasjournal}

\end{document}